\documentclass[prb,twocolumn,floatfix,showpacs,preprintnumbers,amsmath,amssymb,superscriptaddress]{revtex4-1}

\usepackage{graphicx}
\usepackage{graphics}
\usepackage{dcolumn}
\usepackage{bm}

\begin{document}
\DeclareGraphicsExtensions{.pdf,.png,.gif,.jpg}

\title{Field-induced criticality in a gapped quantum magnet with bond disorder.}

 \author{D. H\"uvonen}
 \affiliation{Neutron Scattering and Magnetism, Laboratory for Solid State Physics, ETH Zurich, Switzerland.}
 \author{S. Zhao}
 \affiliation{Neutron Scattering and Magnetism, Laboratory for Solid State Physics, ETH Zurich, Switzerland.}
 \author{M. M\aa nsson}
 \affiliation{Neutron Scattering and Magnetism, Laboratory for Solid State Physics, ETH Zurich, Switzerland.}
 \author{T. Yankova}
 \affiliation{Neutron Scattering and Magnetism, Laboratory for Solid State Physics, ETH Zurich, Switzerland.}
 \affiliation{Chemistry Dept., Lomonosov Moscow State University, Russia.}
 \author{E. Ressouche}
 \affiliation{INAC/SPSMS-MDN, CEA/Grenoble, 17 rue des Martyrs, 38054 Grenoble Cedex 9, France}
 \author{C. Niedermayer}
 \affiliation{Laboratory for Neutron Scattering, Paul Scherrer Institute, CH-5232 Villigen, Switzerland.}
 \author{M. Laver}
 \affiliation{Laboratory for Neutron Scattering, Paul Scherrer Institute, CH-5232 Villigen, Switzerland.}
 \affiliation{Materials Research Division, Risø DTU, Technical University of Denmark, DK-4000 Roskilde, Denmark.}
 \affiliation{Nano Science Center, Niels Bohr Institute, University of Copenhagen, DK-2100 Copenhagen, Denmark.}
 \author{S.N. Gvasaliya}
 \affiliation{Neutron Scattering and Magnetism, Laboratory for Solid State Physics, ETH Zurich, Switzerland.}
 \author{A. Zheludev}
 \homepage{http://www.neutron.ethz.ch/}
 \affiliation{Neutron Scattering and Magnetism, Laboratory for Solid State Physics, ETH Zurich, Switzerland.}
\date{\today}

\begin{abstract}
Neutron diffraction and calorimetric measurements are used to study
the field-induced quantum phase transition in
piperazinium-Cu$_2$(Cl$_{1-x}$Br$_x$)$_6$ ($x=0$, $x=3.5$\% and
$x=7.5$\%), a prototypical quantum antiferromagnet with random
bonds. The critical indexes $\phi$ and $\beta$ are determined.
 The findings contradict some original predictions for
Bose Glass based on the assumption $z=d$, but are consistent with
recent theoretical results implying $z\leq d$.
Inelastic neutron experiments reveal that disorder has a profound effect on the lowest-energy magnetic gap excitation in the system.
\end{abstract}

\pacs{75.10.Jm, 64.70.Tg, 75.40.Cx, 72.15.Rn}

\maketitle

The persistent interest in quantum magnets is driven by their use as
prototypes for the study of quantum critical phenomena. In this
context, much attention has been given to field-induced phase
transitions in gapped spin systems with a singlet ground state, the
so-called quantum spin liquids. Such transitions are described in
terms of a Bose Einstein condensation (BEC) of
magnons.\cite{Giamarchi2008} The specific advantage of magnetic
realizations of BEC lies in the relatively easy experimental access
to- and control of- the key relevant quantities, notably the
effective chemical potential, boson density and BEC order parameter.
In magnetic BEC, the latter correspond to magnetic field,
magnetization and antiferromagnetic order, respectively.

A particular area of recent interest is BEC in the presence of a
spatially random potential. Such disorder leads to a localization of
bosons and the appearance of a novel quantum phase, the so-called
Bose Glass.\cite{Giamarchi1987,Fischer1989,Nohadini2005} Most
importantly, it changes the universality class of the quantum
critical point. While in the disorder-free case it is Mean Field
(MF) like with dynamical critical exponent $z=2$,\cite{Giamarchi2008} very unusual scaling with
$z$ equal to dimensionality ($z=d$) was predicted for disordered systems.\cite{Fischer1989} This
premise was recently challenged,\cite{Weichman2007,Weichman2008}
suggesting that in fact $z\leq d$.\cite{Priyadarshee2006,Meier2011}
Quantum magnetic materials offer a unique opportunity of addressing
this important matter of debate experimentally. A random potential
for magnons can be created by randomizing the strength of magnetic
interactions. This is achieved through a spatially random chemical
modification. Indications of the magnetic  Bose
Glass\cite{Manaka2008,Manaka2009,Hong2010PRBRC} and a modification
of the field-induced ordering
transition\cite{Yamada2011,Yu2011,Wulf2011} have been reported.
However, experiments only added to the confusion regarding critical
indexes and scaling. Recent studies of Tl$_{0.64}$K$_{0.36}$CuCl$_3$
seemed to validate the $z=d$ hypothesis.\cite{Yamada2011} Other
experiments, on the organic compound DTN,\cite{Yu2011} yielded quite
different critical exponents, and were supported by Density Matrix
Renormalization Group (DMRG) calculations. In the present work we
seek to resolve the issue through neutron and
calorimetric studies of the chemically disordered quantum magnets
piperazinium-Cu$_2$(Cl$_{1-x}$Br$_x$)$_6$ (PHCX). We determine the
critical exponents $\phi$ and $\beta$, the latter previously
unmeasured for this type of transition. We find experimental values
that are clearly inconsistent with predictions based on $z=d$.

One rationale for selecting PHCX as a test material is that the
disorder-free Cl-rich ($x=0$) compound, (PHCC) is an exceptionally
well-characterized Heisenberg quantum
magnet.\cite{Stone2001,Stone2006,Stone2007,Stone2006-Nature} It
features a complex quasi two-dimensional spin network of
$S=1/2$ Cu$^{2+}$ ions bridged by Cu-Cl-Cl-Cu superexchange
pathways, see Fig.1 in Ref.\onlinecite{Stone2001}. The exact Hamiltonian for this system remains unknown with 6 to 8 possible superexchange pathways within the Cu$_2$Cl$_6$ layers. However, for the physics discussed here relevant ramification is the spin liquid ground state and the gapped ($\Delta=1$~meV) magnon spectrum with bandwidth of about 1.7~meV. In magnetic
fields exceeding $\mu_0 H_c=7.5$~T, PHCC undergoes a quantum phase
transition to a state with spontaneous antiferromagnetic (AF) order
perpendicular to the direction of applied field. The critical
properties of this transition have been investigated using a variety
of techniques.\cite{Stone2007} The two critical exponents studied in
this work are the order parameter exponent $\beta$ and the so-called
crossover exponent $\phi$. The former defines the field dependence
of the AF ordered moment  at $T\rightarrow 0$:
$|\langle\mathbf{S}_\bot\rangle|\sim (H-H_c)^\beta$, where $H_c \equiv H_c(T=0)$.  The crossover
index $\phi$ defines the phase boundary of the ordered state on the
$H-T$ phase diagram: $T_c(H) \sim (H-H_c)^\phi$ or $T_c(H)^{1/\phi} \sim (H-H_c)$.

Br substitution in PHCX is expected to locally affect the bond
angles in the halogen-mediated superexchange routes, and thereby to
realize the random bond model. This mechanism was previously
exploited in the study of random-bond systems
IPA-CuCl$_3$,\cite{Manaka2008,Hong2010PRBRC} Sul-Cu$_2$Cl$_4$
(Ref.~\onlinecite{Wulf2011}) and DTN.\cite{Yu2011} High quality PHCX
single crystals were grown from solution as described in
Ref.~\onlinecite{Yankova2011}. Single crystal X-ray diffraction
confirmed that for Br content $x$ up  to at least 10\%, the crystal
symmetry remains intact, space group ($P\bar{1}$). The lattice
parameters are $a=7.9691(5)$\AA,$ b=7.0348(5)$\AA,$
 c=6.0836(4)$\AA,$ \alpha=111.083(3)^\circ, \beta=99.947(3)^\circ,
\gamma=81.287(4)^\circ$ for the pure material with
$x=0$,\cite{marcotrigiano1976} and the cell dimensions increase linearly with $x$. 
Here $x$ stands for nominal Br concentration in the starting solution.
Studies on similar compounds have shown a very good correspondence between nominal and final dopant concentrations.\cite{Wulf2011,Yankova2011}
Most of the work reported here was done on perfectly faceted single
crystals with $x=3.5$~\% and $x=7.5$~\%, 
where $a=7.9799(2)$~\AA,$b=7.0406(2)$\AA, $c=6.09090(10)$~\AA 
and  $a=7.9905(3)$~\AA, $b=7.0454(2)$\AA, $c=6.1000(2)$~\AA,
respectively, with the triclinic angles indistinguishable from the pure compound's. 

\begin{figure}[tb]
\includegraphics[width=\columnwidth]{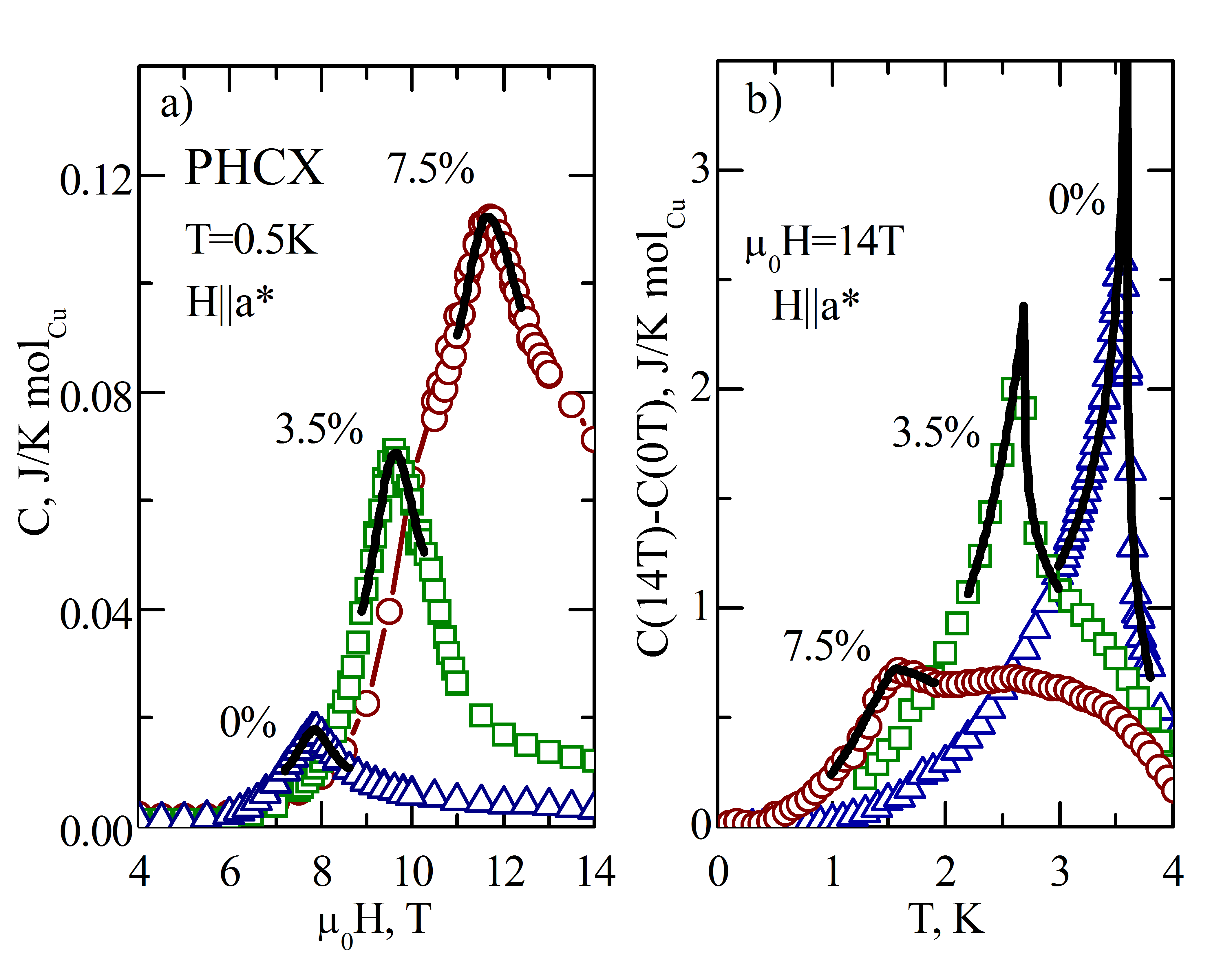}
 \caption{(color
online) Symbols: Heat capacity of PHCX with varying Br concentration $x$
measured (a) as a function of magnetic field at T=0.5K  and (b) as a function of
temperature at $H=14$~T. The solid lines are fits to the data as described
in the text. } \label{fig1}
\end{figure}
\begin{figure}[tb]
\includegraphics[width=\columnwidth]{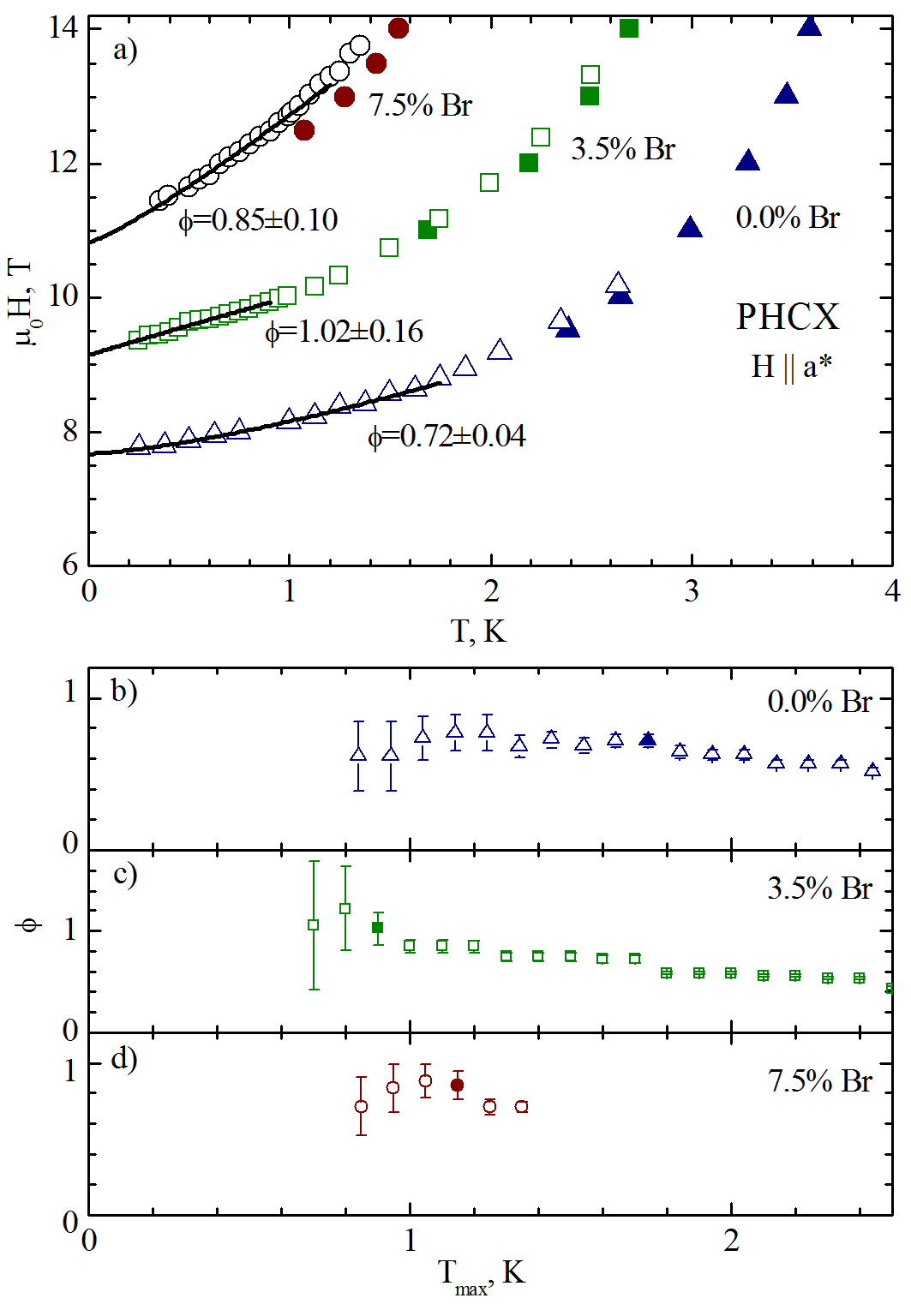}
 \caption{(color online) (a) Phase boundaries of
ordered state for Br-free, 3.5\% and 7.5\% Br substituted PHCX. Open and filled
circles are from $C(H)$ and $C(T)$ measurements, respectively. Solid lines are
a power laws $H-H_c \sim T_c(H)^{1/\phi}$ with critical fields $H_c$ and crossover exponents $\phi$ derived
from the data as described in the text. (b-d) Crossover exponent $\phi$ obtained from
least squares fit as a function of fitting window given by $T < T_{max}$.
Filled symbols show the value quoted in Table \ref{table1} satisfiying our best fit criteria described in the text.}
\label{errors}\label{phasediag}
\end{figure}

The crossover exponent was derived from calorimetric measurements.
For $x=0$, $x=3.5$~\% and $x=7.5$~\%, the data were collected using
a 14~T Quantum Design PPMS with a $^3$He-$^4$He dilution insert,
both as a function of temperature and magnetic field applied along
the $a^{*}$ axis. Typical scans are shown in
Fig.\ref{fig1}a,b (symbols). Sharp lambda-anomalies are apparent in
all cases and mark the ordering transition. 
The apparent broadening at low $T$ for $x=7.5$~\% sample is of instrumental origin. It is caused by differences in the slope of the phase boundary in conjunction with finite temperature resolution.
In the constant-field
scans, the actual transition temperatures $T_c$ were determined by
fitting the data at each field, in the vicinity of the peaks, to an
empirical form of power law with broadening.\cite{hcfitfunc}
A similar approach was used
to obtain the temperature dependence of the critical field $H_c$
from constant-temperature scans. In this case the fit function was a
composite of two half-Lorentzians of non-equal widths and common
maxima. The resulting phase boundaries are plotted in
Fig.~\ref{phasediag} (symbols). Clearly, the introduction of
disorder shifts the transition to higher fields and lower
temperatures.

Any more quantitative analysis has to take into account the experimental
errors. For consistency, we only used the more comprehensive data set obtained
in constant temperature scans (empty symbols in Fig.~\ref{phasediag}). The error
bar on $H_c$ was estimated as the standard deviation of this parameter in the
least squares fits described above and is significantly smaller than the symbol size in the plot.
The dominant error on $T_c$ is of
instrumental origin, and due to the use of the temperature relaxation
measurement method. It is defined by the magnitude of applied heat pulse (3\%
of the starting temperature in our case).

The exponent $\phi$ is obtained by weighted least squares fit of the power law function
to the experimental field-dependence of $T_c$ accounting for experimental errors on both. 
This deceptively simple procedure is actually quite delicate. The fit is performed in
a progressively shrinking data range $T < T_\mathrm{max} \rightarrow 0$.
In Figs.~\ref{errors}b-\ref{errors}d, we show the fitted values of $\phi$ 
for the three materials plotted against $T_\mathrm{max}$.
The range used
for the final fits is to be selected based on two conflicting
requirements. It has to be as narrow as possible, in order to access
the true critical region. This desire is counteracted by the
divergence of confidence intervals at $T_\mathrm{max} \rightarrow 0$. As a
rule of thumb, one can choose the {\it largest} $T_\mathrm{max}$
that gives a value of $\phi$ within the confidence intervals
obtained using any narrower ranges. Further reducing the fitting
range does not produce a statistically significant change in the
result.


\begin{figure}[tb]
\includegraphics[width=\columnwidth]{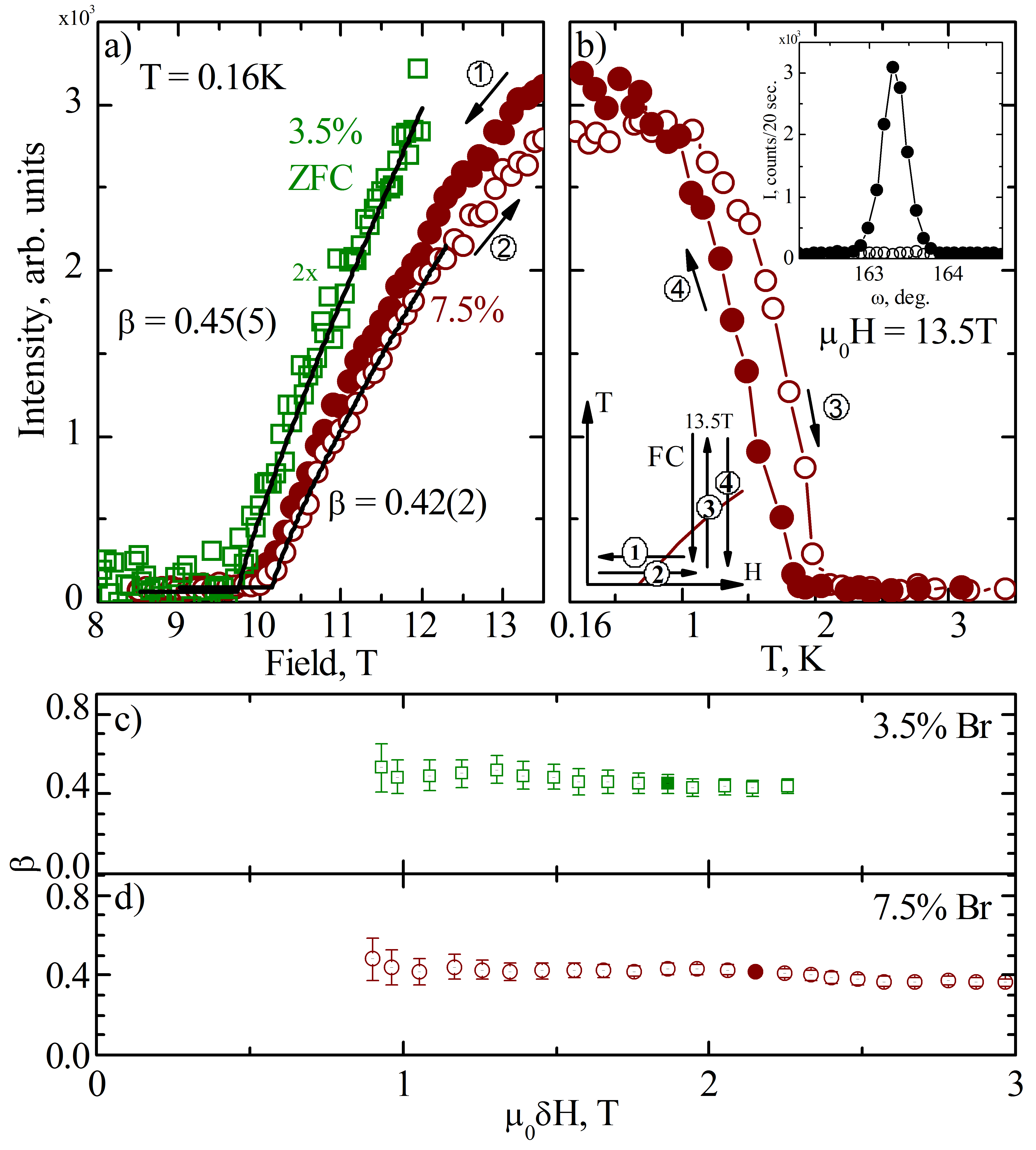}
\caption{(color online) \textit{Upper panels:} Neutron diffraction intensity of
the ($\frac{1}{2} 0 \bar{\frac{1}{2}}$) peak measured in PHCX, a) as a function
of magnetic field at T=160mK, b) as a function of temperature. Open and solid
symbols correspond to zero field cooling and field cooling, respectively.
Squares (circles) correspond to $3.5$\% ($x=7.5$\%) Br content.
Lower left corner: the trajectory in $(H,T)$ space for the measurements on the
$x=7.5$\% material. Inset: rocking curve measured  at $T=0.5$~K (filled) and $T=2$~K (empty circles) in that sample. \textit{Lower panels:} order parameter exponents as obtained by shrinking the fitting interval analysis  of the 3.5\% (panel c)
and 7.5\% Br (panel d) samples data. Filled symbols show the best fit values of $\beta$.}
\label{beta}
\end{figure}


Returning to our case of PHCC, based on the plots in
Fig.~\ref{errors} and the above-mentioned rule of thumb, for the
$x=0$, $x=3.5$\% and $x=7.5$\% materials we have selected fitting
ranges $T_\mathrm{max}=1.75$~K, 0.9~K and 1.15~K, respectively. The resulting
values of $\phi$ and $H_c$ are summarized in Table~\ref{table1}.
Power laws derived from these parameter values are shown as solid
lines up to $T=T_\mathrm{max}$ in Fig.~\ref{phasediag}.

\begin{table}
\begin{tabular}{l|r|r|r} 
\hline\hline \label{criexp}
  & $\mu_0H_c$ T & $\phi$ & $\beta$ \\
\hline PHCX 0.0\% Br &$ 7.67\pm 0.02$ & $0.72\pm 0.04$ & $0.40 \pm 0.01$ [\onlinecite{Stone2007}]
\\
PHCX 3.5\% Br  & $9.14\pm 0.08$ &$1.02\pm 0.16 $ & $ 0.45 \pm 0.05$ \\
PHCX 7.5\% Br  & $10.82\pm 0.14$ &$0.85\pm 0.10$ & $0.42 \pm 0.02$ \\
\hline
DTN 0.8\% Br [\onlinecite{Yu2011}]& $1.07\pm 0.01 $&$1.1\pm 0.2$ & \\
Tl$_{0.64}$K$_{0.36}$CuCl$_3$ [\onlinecite{Yamada2011}]& $2.7\pm 0.6$&$1.7$ & \\
\hline
$z=d$ scaling [\onlinecite{Fischer1989}] & &$\geq 2$ &  \\
QMC [\onlinecite{Priyadarshee2006}] & &$1.54$ & $0.61$ \\
QMC [\onlinecite{Yu2011}] & &$1.06\pm0.09$ & \\
BEC (MF) &  & 0.667  & 0.5 \\
\hline\hline
\end{tabular}
\caption{Experimental estimates for the critical field $H_c$ in $\mathbf{H}||\mathbf{a^*}$ orientation and the
critical exponents $\phi$ (from calorimetric data) and $\beta$ (from
neutron diffraction) of the field-induced quantum phase transition
in PHCX in comparison with recent measurements on similar systems
and theoretical results, whose relevance is discussed in the
text.\label{table1}}
\end{table}

The behavior of the magnetic BEC order parameter  was studied in
neutron diffraction experiments. For the $x=3.5$\% and $x=7.5$\%
samples these were carried out at the CRG-CEA D23 diffractometer at ILL and the 
RITA-II 3-axis spectrometer at PSI, respectively. In the two
experiments, we used Cu monochromator to produce
neutrons of incident energies 50~meV at ILL and pyrolitic graphite for 4.6~meV neutrons with Be filter in front of the analyzer at PSI. 
Sample environment
was in all cases a split coil cryomagnet with a $^3$He-$^4$He
dilution cryostat. The magnetic field was applied along $\mathbf{a^*}$ in the first and along the
$\mathbf{b}$ crystallographic axis in the second experiment. As previously reported for
PHCC,\cite{Stone2007} long range order in the BEC phase is marked by
the appearance of new magnetic Bragg reflections half-integer
reciprocal-lattice points $(h+1/2, 0, l+1/2)$. The inset in
Fig.\ref{beta}b shows typical scans across the $(0.5,0,-0.5)$
position measured in the $x=7.5$\% sample at $T=0.5$~K in the BEC
phase, and the background collected at $T=2$~K in the paramagnetic
state. To within experimental error, all magnetic reflections are
resolution-limited.

The critical index $\beta$ was determined  from measurements of the
$(0.5,0,-0.5)$ Bragg peak intensity in zero field cooled (ZFC)
samples. These data were taken upon increasing the applied field at
$T=160$~mK and are plotted in Fig.\ref{beta}a. $\beta$ was then
determined in power law fits, assuming that peak intensity scales as
as the square of the order parameter. As for $\phi$, the fits for
$\beta$ were performed for a series of shrinking fitting intervals, of magnetic field
$\delta H = H- H_c$  in this case. The fit results and standard deviation confidence
intervals are plotted against $\delta H$ in Figs.~\ref{beta}c,d.
Following the reasoning described earlier, in our final analysis we
have selected $\mu_0\delta H=1.86$~T and 2.15~T, for $x=3.5$\% and $x=7.5$\%,
respectively. The resulting exponents are summarized in
Table~\ref{table1}. In order to properly characterize the quantum
critical point, one has to know $\beta$ in the limit $T\rightarrow
0$. Previous studies of PHCC suggest that $\beta$ remains
practically constant for $T<0.3$~K.\cite{Stone2007} We have
additionally verified this behavior for our 3.5\% PHCX sample. Thus,
the 160~mK measurements are good estimates of the zero-temperature
values. Comparing our results with those reported in
Ref.~\onlinecite{Stone2007}, we conclude that disorder does not
significantly change the criticality of the order parameter.

\begin{figure}[tb]
\includegraphics[width=\columnwidth]{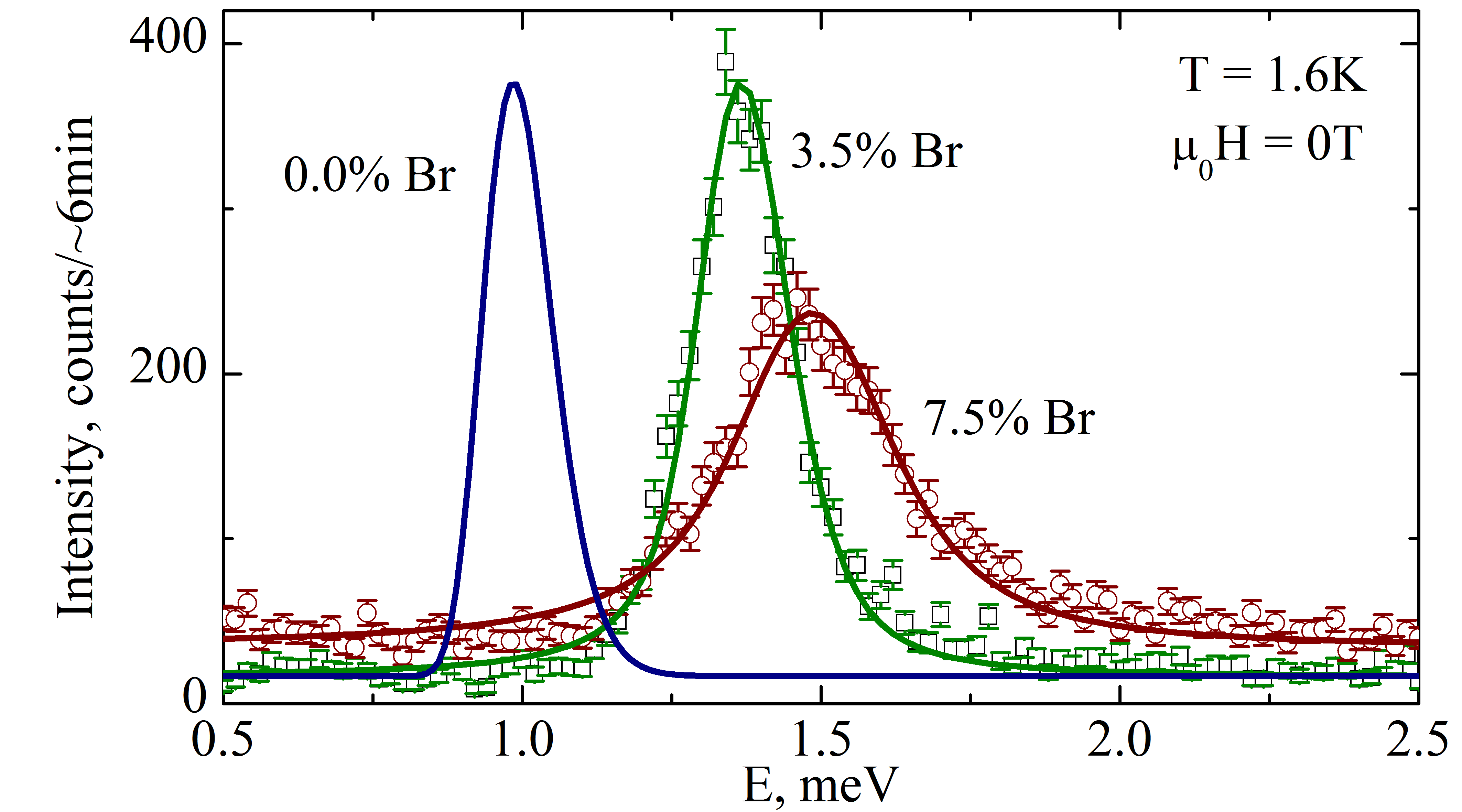}
\caption{(color online) Inelastic neutron scattering intensity at (0.5,0.5,-0.5) in energy transfer range from 0.5 to 2.5 meV for $3.5$\% (green squares) and $x=7.5$\% Br (red circles) content samples. Corresponding solid lines are the fitted Lorentzian energy profiles convoluted with the 4-dimensional instrumental resolution function. For 0.0\% Br sample resolution convoluted simulation is shown for comparison using parameters from Ref.\onlinecite{Stone2001}}
\label{ins}
\end{figure}

This is not to say that the effect of disorder is negligible.
Inelastic neutron scattering experiments performed at zero field on the TASP 3-axis spectrometer\cite{Semadeni01} at PSI using 3.5meV incident-energy neutrons and Be filter
 reveal that while the magnons are sharp in
PHCC,\cite{Stone2007} they acquire a considerable intrinsic width in
disordered samples. 
Typical constant-q scans at the AF zone-center (0.5,0.5,-0.5) in the x=3.5\% and x=7.5\% samples are shown in Fig. \ref{ins} by the open symbols. 
The solid lines are fits of Lorentzian energy profiles convoluted with the 4-dimensional instrumental resolution function \cite{Popovici75}. 
For x=0\% the line is a simulation assuming that peak shape is determined solely by resolution.
The observed increase of gap energy with Br substitution is fully consistent with the observed increase of critical field and may be attributed to a change in the disorder-averaged exchange constants.
However, the large intrinsic energy width $\Gamma=0.1$~meV and 0.25~meV for x=3.5\% and x=7.5\% samples, respectively, is entirely a disorder-induced effect \cite{danphccins}. 

Our present diffraction experiments show that
disorder also affects the behavior of the order parameter deep within
the high-field phase. In the 7.5\% material the intensity of the
magnetic reflection exhibits a history dependence. This is
illustrated in Fig.~\ref{beta}a,b. Here we show data collected along
a trajectory illustrated in the inset. No such behavior was reported
for the disorder-free material, but a similar, and much stronger
history effect was previously studied in a related system
IPA-(Cu$_{1-x}$Br$_x$)Cl$_3$.\cite{Hong2010PRBRC} 
Related phase transition scenario that could lead to  
hysteretic behavior was discussed in Ref.~\onlinecite{Yu2010}.

Theoretical and numerical predictions for the two measured critical
exponents are shown in Table~\ref{table1} for a direct comparison
with our experiments. In the absence of disorder, one expects MF
values\cite{Stone2007} $\beta=1/2$ and $\phi=2/3$, in reasonable agreement with the data for PHCC. Assuming $z=d$ the disordered case gives much
larger value for the crossover exponent $\phi\geq 2$.\cite{phiscaling}
This prediction is {\it clearly inconsistent with our
experiments}. More recent numerical
work\cite{Priyadarshee2006,Yu2011,Meier2011} suggested smaller
values of the exponent. Those from
Refs.~\onlinecite{Priyadarshee2006,Meier2011} are less relevant to
3-dimensional ordering in PHCX, as they are obtained for a
2-dimensional model. On the other hand, the QMC calculations of
Ref.~\onlinecite{Yu2011} should be directly comparable to our data.
The corresponding estimates of $\phi$ are in a rather good agreement
with our experiments, as well as with those on the Br-substituted
DTN compound.\cite{Yu2011} A relevant theoretical study of $\beta$
is presently lacking. The claimed
observation\cite{Yamada2011} of values consistent with $z=d$ can be 
due to overinterpretation of limited dataset.
Evidence is mounting for the breakdown of the $z=d$ hypothesis in
the Bose Glass to BEC quantum phase transition.

This work is partially supported by the Swiss National Fund under
project 2-77060-11 and through Project 6 of MANEP.
ML acknowledges support from DanScat.
 We thank Dr. V. Glazkov for his involvement in the early stages of this project.


\begin{thebibliography}{25}%
\makeatletter
\providecommand \@ifxundefined [1]{%
 \@ifx{#1\undefined}
}%
\providecommand \@ifnum [1]{%
 \ifnum #1\expandafter \@firstoftwo
 \else \expandafter \@secondoftwo
 \fi
}%
\providecommand \@ifx [1]{%
 \ifx #1\expandafter \@firstoftwo
 \else \expandafter \@secondoftwo
 \fi
}%
\providecommand \natexlab [1]{#1}%
\providecommand \enquote  [1]{``#1''}%
\providecommand \bibnamefont  [1]{#1}%
\providecommand \bibfnamefont [1]{#1}%
\providecommand \citenamefont [1]{#1}%
\providecommand \href@noop [0]{\@secondoftwo}%
\providecommand \href [0]{\begingroup \@sanitize@url \@href}%
\providecommand \@href[1]{\@@startlink{#1}\@@href}%
\providecommand \@@href[1]{\endgroup#1\@@endlink}%
\providecommand \@sanitize@url [0]{\catcode `\\12\catcode `\$12\catcode
  `\&12\catcode `\#12\catcode `\^12\catcode `\_12\catcode `\%12\relax}%
\providecommand \@@startlink[1]{}%
\providecommand \@@endlink[0]{}%
\providecommand \url  [0]{\begingroup\@sanitize@url \@url }%
\providecommand \@url [1]{\endgroup\@href {#1}{\urlprefix }}%
\providecommand \urlprefix  [0]{URL }%
\providecommand \Eprint [0]{\href }%
\providecommand \doibase [0]{http://dx.doi.org/}%
\providecommand \selectlanguage [0]{\@gobble}%
\providecommand \bibinfo  [0]{\@secondoftwo}%
\providecommand \bibfield  [0]{\@secondoftwo}%
\providecommand \translation [1]{[#1]}%
\providecommand \BibitemOpen [0]{}%
\providecommand \bibitemStop [0]{}%
\providecommand \bibitemNoStop [0]{.\EOS\space}%
\providecommand \EOS [0]{\spacefactor3000\relax}%
\providecommand \BibitemShut  [1]{\csname bibitem#1\endcsname}%
\let\auto@bib@innerbib\@empty
\bibitem [{\citenamefont {Giamarchi}\ \emph {et~al.}(2008)\citenamefont
  {Giamarchi}, \citenamefont {Ruegg},\ and\ \citenamefont
  {Tchernyshev}}]{Giamarchi2008}%
  \BibitemOpen
  \bibfield  {author} {\bibinfo {author} {\bibfnamefont {T.}~\bibnamefont
  {Giamarchi}}, \bibinfo {author} {\bibfnamefont {C.}~\bibnamefont {Ruegg}}, \
  and\ \bibinfo {author} {\bibfnamefont {O.}~\bibnamefont {Tchernyshev}},\
  }\href@noop {} {\bibfield  {journal} {\bibinfo  {journal} {Nature Physics}\
  }\textbf {\bibinfo {volume} {4}},\ \bibinfo {pages} {198} (\bibinfo {year}
  {2008})}\BibitemShut {NoStop}%
\bibitem [{\citenamefont {Giamarchi}\ and\ \citenamefont
  {Schulz}(1987)}]{Giamarchi1987}%
  \BibitemOpen
  \bibfield  {author} {\bibinfo {author} {\bibfnamefont {T.}~\bibnamefont
  {Giamarchi}}\ and\ \bibinfo {author} {\bibfnamefont {H.~J.}\ \bibnamefont
  {Schulz}},\ }\href@noop {} {\bibfield  {journal} {\bibinfo  {journal}
  {Europhys. Lett.}\ }\textbf {\bibinfo {volume} {3}},\ \bibinfo {pages} {1287}
  (\bibinfo {year} {1987})}\BibitemShut {NoStop}%
\bibitem [{\citenamefont {Fischer}\ \emph {et~al.}(1989)\citenamefont
  {Fischer}, \citenamefont {Weichman}, \citenamefont {Grinstein},\ and\
  \citenamefont {Fischer}}]{Fischer1989}%
  \BibitemOpen
  \bibfield  {author} {\bibinfo {author} {\bibfnamefont {M.~P.~A.}\
  \bibnamefont {Fisher}}, \bibinfo {author} {\bibfnamefont {P.~B.}\
  \bibnamefont {Weichman}}, \bibinfo {author} {\bibfnamefont {G.}~\bibnamefont
  {Grinstein}}, \ and\ \bibinfo {author} {\bibfnamefont {D.~S.}\ \bibnamefont
  {Fisher}},\ }\href@noop {} {\bibfield  {journal} {\bibinfo  {journal} {Phys.
  Rev. B}\ }\textbf {\bibinfo {volume} {40}},\ \bibinfo {pages} {546} (\bibinfo
  {year} {1989})}\BibitemShut {NoStop}%
\bibitem [{\citenamefont {Nohadini}\ \emph {et~al.}(2005)\citenamefont
  {Nohadini}, \citenamefont {Wessel},\ and\ \citenamefont
  {Haas}}]{Nohadini2005}%
  \BibitemOpen
  \bibfield  {author} {\bibinfo {author} {\bibfnamefont {O.}~\bibnamefont
  {Nohadani}}, \bibinfo {author} {\bibfnamefont {S.}~\bibnamefont {Wessel}}, \
  and\ \bibinfo {author} {\bibfnamefont {S.}~\bibnamefont {Haas}},\ }\href@noop
  {} {\bibfield  {journal} {\bibinfo  {journal} {Phys. Rev. Lett.}\ }\textbf
  {\bibinfo {volume} {95}},\ \bibinfo {pages} {227201} (\bibinfo {year}
  {2005})}\BibitemShut {NoStop}%
\bibitem [{\citenamefont {Weichman}\ and\ \citenamefont
  {Mukhopadhyay}(2007)}]{Weichman2007}%
  \BibitemOpen
  \bibfield  {author} {\bibinfo {author} {\bibfnamefont {P.~B.}\ \bibnamefont
  {Weichman}}\ and\ \bibinfo {author} {\bibfnamefont {R.}~\bibnamefont
  {Mukhopadhyay}},\ }\href@noop {} {\bibfield  {journal} {\bibinfo  {journal}
  {Physical Review Letters}\ }\textbf {\bibinfo {volume} {98}},\ \bibinfo
  {pages} {245701} (\bibinfo {year} {2007})}\BibitemShut {NoStop}%
\bibitem [{\citenamefont {Weichman}\ and\ \citenamefont
  {Mukhopadhyay}(2008)}]{Weichman2008}%
  \BibitemOpen
  \bibfield  {author} {\bibinfo {author} {\bibfnamefont {P.~B.}\ \bibnamefont
  {Weichman}}\ and\ \bibinfo {author} {\bibfnamefont {R.}~\bibnamefont
  {Mukhopadhyay}},\ }\href@noop {} {\bibfield  {journal} {\bibinfo  {journal}
  {Physical Review B}\ }\textbf {\bibinfo {volume} {77}},\ \bibinfo {pages}
  {214516} (\bibinfo {year} {2008})}\BibitemShut {NoStop}%
\bibitem [{\citenamefont {Priyadarshee}\ \emph {et~al.}(2006)\citenamefont
  {Priyadarshee}, \citenamefont {Chandrasekharan}, \citenamefont {Lee},\ and\
  \citenamefont {Baranger}}]{Priyadarshee2006}%
  \BibitemOpen
  \bibfield  {author} {\bibinfo {author} {\bibfnamefont {A.}~\bibnamefont
  {Priyadarshee}}, \bibinfo {author} {\bibfnamefont {S.}~\bibnamefont
  {Chandrasekharan}}, \bibinfo {author} {\bibfnamefont {J.-W.}\ \bibnamefont
  {Lee}}, \ and\ \bibinfo {author} {\bibfnamefont {H.~U.}\ \bibnamefont
  {Baranger}},\ }\href@noop {} {\bibfield  {journal} {\bibinfo  {journal}
  {Physical Review Letters}\ }\textbf {\bibinfo {volume} {97}},\ \bibinfo
  {pages} {115703} (\bibinfo {year} {2006})}\BibitemShut {NoStop}%
\bibitem [{Mei()}]{Meier2011}%
  \BibitemOpen
  \href@noop {} {}\bibinfo {note} {H. Meier and M. Wallin,
  arXiv:1109.3022v2}\BibitemShut {NoStop}%
\bibitem [{\citenamefont {Manaka}\ \emph {et~al.}(2008)\citenamefont {Manaka},
  \citenamefont {Kolomiets},\ and\ \citenamefont {Goto}}]{Manaka2008}%
  \BibitemOpen
  \bibfield  {author} {\bibinfo {author} {\bibfnamefont {H.}~\bibnamefont
  {Manaka}}, \bibinfo {author} {\bibfnamefont {A.~V.}\ \bibnamefont
  {Kolomiets}}, \ and\ \bibinfo {author} {\bibfnamefont {T.}~\bibnamefont
  {Goto}},\ }\href {\doibase 10.1103/PhysRevLett.101.077204} {\bibfield
  {journal} {\bibinfo  {journal} {Phys. Rev. Lett.}\ }\textbf {\bibinfo
  {volume} {101}},\ \bibinfo {pages} {077204} (\bibinfo {year}
  {2008})}\BibitemShut {NoStop}%
\bibitem [{\citenamefont {Manaka}\ \emph {et~al.}(2009)\citenamefont {Manaka},
  \citenamefont {Katori}, \citenamefont {Kolomiets},\ and\ \citenamefont
  {Goto}}]{Manaka2009}%
  \BibitemOpen
  \bibfield  {author} {\bibinfo {author} {\bibfnamefont {H.}~\bibnamefont
  {Manaka}}, \bibinfo {author} {\bibfnamefont {H.~A.}\ \bibnamefont {Katori}},
  \bibinfo {author} {\bibfnamefont {O.~V.}\ \bibnamefont {Kolomiets}}, \ and\
  \bibinfo {author} {\bibfnamefont {T.}~\bibnamefont {Goto}},\ }\href {\doibase
  10.1103/PhysRevB.79.092401} {\bibfield  {journal} {\bibinfo  {journal} {Phys.
  Rev. B}\ }\textbf {\bibinfo {volume} {79}},\ \bibinfo {pages} {092401}
  (\bibinfo {year} {2009})}\BibitemShut {NoStop}%
\bibitem [{\citenamefont {Hong}\ \emph {et~al.}(2010)\citenamefont {Hong},
  \citenamefont {Zheludev}, \citenamefont {Manaka},\ and\ \citenamefont
  {Regnault}}]{Hong2010PRBRC}%
  \BibitemOpen
  \bibfield  {author} {\bibinfo {author} {\bibfnamefont {T.}~\bibnamefont
  {Hong}}, \bibinfo {author} {\bibfnamefont {A.}~\bibnamefont {Zheludev}},
  \bibinfo {author} {\bibfnamefont {H.}~\bibnamefont {Manaka}}, \ and\ \bibinfo
  {author} {\bibfnamefont {L.-P.}\ \bibnamefont {Regnault}},\ }\href {\doibase
  10.1103/PhysRevB.81.060410} {\bibfield  {journal} {\bibinfo  {journal} {Phys.
  Rev. B}\ }\textbf {\bibinfo {volume} {81}},\ \bibinfo {pages} {060410}
  (\bibinfo {year} {2010})}\BibitemShut {NoStop}%
\bibitem [{\citenamefont {Yamada}\ \emph {et~al.}(2011)\citenamefont {Yamada},
  \citenamefont {Tanaka}, \citenamefont {Ono},\ and\ \citenamefont
  {Nojiri}}]{Yamada2011}%
  \BibitemOpen
  \bibfield  {author} {\bibinfo {author} {\bibfnamefont {F.}~\bibnamefont
  {Yamada}}, \bibinfo {author} {\bibfnamefont {H.}~\bibnamefont {Tanaka}},
  \bibinfo {author} {\bibfnamefont {T.}~\bibnamefont {Ono}}, \ and\ \bibinfo
  {author} {\bibfnamefont {H.}~\bibnamefont {Nojiri}},\ }\href {\doibase
  10.1103/PhysRevB.83.020409} {\bibfield  {journal} {\bibinfo  {journal} {Phys.
  Rev. B}\ }\textbf {\bibinfo {volume} {83}},\ \bibinfo {pages} {020409}
  (\bibinfo {year} {2011})}\BibitemShut {NoStop}%
\bibitem [{Yu2()}]{Yu2011}%
  \BibitemOpen
  \href@noop {} {}\bibinfo {note} {R. Yu, L. Yin, N. S. Sullivan, J. S. Xia, C.
  Huan, A. Paduan-Filho, N. F. Oliveira Jr., S. Haas, A. Steppke, C. F. Miclea,
  F. Weickert, R. Movshovich, E.-D. Mun, V. S. Zapf and T. Roscilde,
  arXiv:1109.4403v2.}\BibitemShut {Stop}%
\bibitem [{\citenamefont {Wulf}\ \emph {et~al.}(2011)\citenamefont {Wulf},
  \citenamefont {M\"uhlbauer}, \citenamefont {Yankova},\ and\ \citenamefont
  {Zheludev}}]{Wulf2011}%
  \BibitemOpen
  \bibfield  {author} {\bibinfo {author} {\bibfnamefont {E.}~\bibnamefont
  {Wulf}}, \bibinfo {author} {\bibfnamefont {S.}~\bibnamefont {M\"uhlbauer}},
  \bibinfo {author} {\bibfnamefont {T.}~\bibnamefont {Yankova}}, \ and\
  \bibinfo {author} {\bibfnamefont {A.}~\bibnamefont {Zheludev}},\ }\href
  {\doibase 10.1103/PhysRevB.84.174414} {\bibfield  {journal} {\bibinfo
  {journal} {Phys. Rev. B}\ }\textbf {\bibinfo {volume} {84}},\ \bibinfo
  {pages} {174414} (\bibinfo {year} {2011})}\BibitemShut {NoStop}%
\bibitem [{\citenamefont {Stone}\ \emph {et~al.}(2001)\citenamefont {Stone},
  \citenamefont {Zaliznyak}, \citenamefont {Reich},\ and\ \citenamefont
  {Broholm}}]{Stone2001}%
  \BibitemOpen
  \bibfield  {author} {\bibinfo {author} {\bibfnamefont {M.~B.}\ \bibnamefont
  {Stone}}, \bibinfo {author} {\bibfnamefont {I.}~\bibnamefont {Zaliznyak}},
  \bibinfo {author} {\bibfnamefont {D.~H.}\ \bibnamefont {Reich}}, \ and\
  \bibinfo {author} {\bibfnamefont {C.}~\bibnamefont {Broholm}},\ }\href
  {\doibase 10.1103/PhysRevB.64.144405} {\bibfield  {journal} {\bibinfo
  {journal} {Phys. Rev. B}\ }\textbf {\bibinfo {volume} {64}},\ \bibinfo
  {pages} {144405} (\bibinfo {year} {2001})}\BibitemShut {NoStop}%
\bibitem [{\citenamefont {Stone}\ \emph
  {et~al.}(2006{\natexlab{a}})\citenamefont {Stone}, \citenamefont {Broholm},
  \citenamefont {Reich}, \citenamefont {Tchernyshyov}, \citenamefont
  {Vorderwisch},\ and\ \citenamefont {Harrison}}]{Stone2006}%
  \BibitemOpen
  \bibfield  {author} {\bibinfo {author} {\bibfnamefont {M.~B.}\ \bibnamefont
  {Stone}}, \bibinfo {author} {\bibfnamefont {C.}~\bibnamefont {Broholm}},
  \bibinfo {author} {\bibfnamefont {D.~H.}\ \bibnamefont {Reich}}, \bibinfo
  {author} {\bibfnamefont {O.}~\bibnamefont {Tchernyshyov}}, \bibinfo {author}
  {\bibfnamefont {P.}~\bibnamefont {Vorderwisch}}, \ and\ \bibinfo {author}
  {\bibfnamefont {N.}~\bibnamefont {Harrison}},\ }\href {\doibase
  10.1103/PhysRevLett.96.257203} {\bibfield  {journal} {\bibinfo  {journal}
  {Phys. Rev. Lett.}\ }\textbf {\bibinfo {volume} {96}},\ \bibinfo {pages}
  {257203} (\bibinfo {year} {2006}{\natexlab{a}})}\BibitemShut {NoStop}%
\bibitem [{\citenamefont {Stone}\ \emph {et~al.}(2007)\citenamefont {Stone},
  \citenamefont {Broholm}, \citenamefont {Reich}, \citenamefont {Schiffer},
  \citenamefont {Tchernyshyov}, \citenamefont {Vorderwisch},\ and\
  \citenamefont {Harrison}}]{Stone2007}%
  \BibitemOpen
  \bibfield  {author} {\bibinfo {author} {\bibfnamefont {M.~B.}\ \bibnamefont
  {Stone}}, \bibinfo {author} {\bibfnamefont {C.}~\bibnamefont {Broholm}},
  \bibinfo {author} {\bibfnamefont {D.~H.}\ \bibnamefont {Reich}}, \bibinfo
  {author} {\bibfnamefont {P.}~\bibnamefont {Schiffer}}, \bibinfo {author}
  {\bibfnamefont {O.}~\bibnamefont {Tchernyshyov}}, \bibinfo {author}
  {\bibfnamefont {P.}~\bibnamefont {Vorderwisch}}, \ and\ \bibinfo {author}
  {\bibfnamefont {N.}~\bibnamefont {Harrison}},\ }\href
  {http://stacks.iop.org/1367-2630/9/i=2/a=031} {\bibfield  {journal} {\bibinfo
   {journal} {New Journal of Physics}\ }\textbf {\bibinfo {volume} {9}},\
  \bibinfo {pages} {31} (\bibinfo {year} {2007})}\BibitemShut {NoStop}%
\bibitem [{\citenamefont {Stone}\ \emph
  {et~al.}(2006{\natexlab{b}})\citenamefont {Stone}, \citenamefont {Zaliznyak},
  \citenamefont {Hong}, \citenamefont {Broholm},\ and\ \citenamefont
  {Reich}}]{Stone2006-Nature}%
  \BibitemOpen
  \bibfield  {author} {\bibinfo {author} {\bibfnamefont {M.~B.}\ \bibnamefont
  {Stone}}, \bibinfo {author} {\bibfnamefont {I.~A.}\ \bibnamefont
  {Zaliznyak}}, \bibinfo {author} {\bibfnamefont {T.}~\bibnamefont {Hong}},
  \bibinfo {author} {\bibfnamefont {C.~L.}\ \bibnamefont {Broholm}}, \ and\
  \bibinfo {author} {\bibfnamefont {D.~H.}\ \bibnamefont {Reich}},\ }\href@noop
  {} {\bibfield  {journal} {\bibinfo  {journal} {Nature}\ }\textbf {\bibinfo
  {volume} {440}},\ \bibinfo {pages} {187} (\bibinfo {year}
  {2006}{\natexlab{b}})}\BibitemShut {NoStop}%
\bibitem [{Yan()}]{Yankova2011}%
  \BibitemOpen
  \href@noop {} {}\bibinfo {note} {T. Yankova, D. H{\"u}vonen, S. Muehlbauer, D.
  Schmidiger, E. Wulf, S. Zhao, A. Zheludev, T. Hong, V. O. Garlea, R.
  Custelcean, G. Ehlers, arXiv:1110.6375v1.}\BibitemShut {Stop}%
\bibitem [{\citenamefont {Marcotrigiano}\ \emph {et~al.}(1976)\citenamefont
  {Marcotrigiano}, \citenamefont {Menabue},\ and\ \citenamefont
  {Pellacani}}]{marcotrigiano1976}%
  \BibitemOpen
  \bibfield  {author} {\bibinfo {author} {\bibfnamefont {G.}~\bibnamefont
  {Marcotrigiano}}, \bibinfo {author} {\bibfnamefont {L.}~\bibnamefont
  {Menabue}}, \ and\ \bibinfo {author} {\bibfnamefont {G.~C.}\ \bibnamefont
  {Pellacani}},\ }\href {\doibase 10.1021/ic50164a002} {\bibfield  {journal}
  {\bibinfo  {journal} {Inorganic Chemistry}\ }\textbf {\bibinfo {volume}
  {15}},\ \bibinfo {pages} {2333} (\bibinfo {year} {1976})}\BibitemShut
  {NoStop}%
\bibitem [{hcf()}]{hcfitfunc}%
  \BibitemOpen
  \href@noop {} {}\bibinfo {note} {$C(T) =
  b-a_{l,r}\gamma^{-2}-a_{l,r}\left(|T-T_c|^{\alpha_{l,r}}+\gamma^{2}\right)^{%
-1}$ where $\gamma$ is a broadening parameter, $b$ is background, $a_{l,r}$ and
  $\alpha_{l,r}$ are scaling constants and scaling exponents on the left and
  right side of $T_c$ respectively.}\BibitemShut {Stop}%
\bibitem [{\citenamefont {Semadeni}\ \emph {et~al.}(2001)\citenamefont
  {Semadeni}, \citenamefont {Roessli},\ and\ \citenamefont
  {Boni}}]{Semadeni01}%
  \BibitemOpen
  \bibfield  {author} {\bibinfo {author} {\bibfnamefont {F.}~\bibnamefont
  {Semadeni}}, \bibinfo {author} {\bibfnamefont {B.}~\bibnamefont {Roessli}}, \
  and\ \bibinfo {author} {\bibfnamefont {P.}~\bibnamefont {Boni}},\ }\href@noop
  {} {\bibfield  {journal} {\bibinfo  {journal} {Physica B}\ }\textbf {\bibinfo
  {volume} {297}},\ \bibinfo {pages} {152} (\bibinfo {year}
  {2001})}\BibitemShut {NoStop}%
\bibitem [{\citenamefont {Popovici}(1975)}]{Popovici75}%
  \BibitemOpen
  \bibfield  {author} {\bibinfo {author} {\bibfnamefont {M.}~\bibnamefont
  {Popovici}},\ }\href@noop {} {\bibfield  {journal} {\bibinfo  {journal} {Acta
  Cryst.}\ }\textbf {\bibinfo {volume} {A31}},\ \bibinfo {pages} {507}
  (\bibinfo {year} {1975})}\BibitemShut {NoStop}%
\bibitem [{dan()}]{danphccins}%
  \BibitemOpen
  \href@noop {} {}\bibinfo {note} {A full account of the inelastic neutron
  scattering experiments are beyond the scope of the present paper that is
  focused on the field-induced transitions. It will be made
  elsewhere.}\BibitemShut {Stop}%
\bibitem [{dan()}]{phiscaling}%
  \BibitemOpen
  \href@noop {} {}\bibinfo {note} {Lower limit on $\phi$ follows from Eqs. 5.8 and 5.10 in Ref. 3 as follows $T_c \propto {[\rho_s(0)]}^x$ and $\rho_s(0) \propto {(\rho-\rho_c)}^{\zeta}$ with $x=z/(d+z-2)$ and $\zeta=\nu(d+z-2)$. Since magnetization and boson density are proportianal $(\rho-\rho_c) \propto (H-H_c)$ one obtains $T_c \propto {(H-H_c)}^{x\zeta} = {(H-H_c)}^{z\nu}$. Since $\nu\geq 2/d$ [Eq. 3.24 in Ref. 3] for all dimensions in the presence of disorder and $z=d$ we obtain dimension invariant $\phi\geq 2$.}\BibitemShut {Stop}%
\bibitem [{\citenamefont {Yu}\ \emph {et~al.}(2010)\citenamefont {Yu},
  \citenamefont {Haas},\ and\ \citenamefont {Roscilde}}]{Yu2010}%
  \BibitemOpen
  \bibfield  {author} {\bibinfo {author} {\bibfnamefont {R.}~\bibnamefont
  {Yu}}, \bibinfo {author} {\bibfnamefont {S.}~\bibnamefont {Haas}}, \ and\
  \bibinfo {author} {\bibfnamefont {T.}~\bibnamefont {Roscilde}},\ }\href@noop
  {} {\bibfield  {journal} {\bibinfo  {journal} {Europhys. Lett.}\ }\textbf
  {\bibinfo {volume} {89}},\ \bibinfo {pages} {10009} (\bibinfo {year}
  {2010})}\BibitemShut {NoStop}%
\end{thebibliography}

%

\end{document}